\documentclass{article}
\usepackage[T1]{fontenc}

\makeatletter

\newcommand{\LyX}{L\kern-.1667em\lower.25em\hbox{Y}\kern-.125emX\@}

\makeatother

\begin{document}

\title{Incommensurate structure factor in a hole-doped spin-1 system }

\author{Indrani Bose and Emily Chattopadhyay}

\maketitle
\begin{center}

\textbf{Department of Physics}

\textbf{Bose Institute}

\textbf{93/1, A.P.C. Road, Calcutta-700009}

\textbf{India}

\end{center}

\begin{abstract}
The nickelate compound \( Y_{2}BaNiO_{5} \) is a spin-1 Haldane-gap antiferromagnet.
The compound is doped with holes on replacing the off-chain \( Y^{3+} \) ions
by \( Ca^{2+} \) ions. Inelastic neutron scattering (INS) experiments reveal
the existence of sub-gap states on doping. A recent INS experiment provides
evidence for an incommensurate double-peaked structure factor S(q) corresponding
to the sub-gap states. In this paper, we formulate a microscopic theory for
the origin of the incommensurate peaks.

P.A.C.S. Nos.: 75.10 Jm, 71.27.+a
\end{abstract}

\section*{1.Introduction}

Hole-doped quantum spin systems exhibit a variety of novel phenomena as a function of the dopant concentration. The cuprates, which become high-temperature superconductors on doping, are the prime examples of such systems. The cuprates have a rich T versus x phase diagram, where T is the temperature and x the dopant concentration\( ^{1-3} \). In the undoped state, the compounds exhibit antiferromagnetic (AFM) long range order (LRO) below the N\'eel temperature \( T_{N} \). The dominant electronic and magnetic properties of the cuprates are associated with the copper-oxide (\( CuO_{2} \)) planes. On the introduction of a few percent of holes, the AFM LRO is rapidly destroyed leaving behind a spin-disordered state. The hole-doped \( CuO_{2} \) plane is an example of a quantum spin liquid (QSL) in 2d. Simpler AFM doped spin systems include the two-chain ladder compounds \( LaCuO_{2.5} \)\( ^{4} \) and \( Sr_{14-x}Ca_{x}Cu_{24}O_{41} \)\( ^{5} \) and the spin-1, linear chain AFM compound \( Y_{2-x}Ca_{x}BaNiO_{5} \)\( ^{6} \). These quasi-1d systems exhibit phenomena some of which are similar to those observed in the cuprate systems. For example, the doped ladder compound \( Sr_{14-x}Ca_{x}Cu_{24}O_{41} \)\( ^{5,7} \) becomes a superconductor under high pressure   (\( T_{c}\sim 12K \) at a pressure of 3 Gpa). As in the cuprate systems, holes form bound pairs in the superconducting state.

The doped spin-1 compound \( Y_{2-x}Ca_{x}BaNiO_{5} \) provides an example
of a QSL in 1d. The parent compound \( Y_{2}BaNiO_{5} \) is a charge transfer
insulator containing \( Ni^{2+} \) (S = 1) chains. The ground state of the
system is spin-disordered and the spin excitation spectrum is separated from
the ground state by an energy gap, the so-called Haldane gap\( ^{8} \). The
compound is doped with holes on replacing the off-chain \( Y^{3+} \) ions by
\( Ca^{2+} \) ions. The holes mostly appear in oxygen orbitals along the NiO
chains. No metal-insulator transition takes place but the DC-resistivity \( \rho _{DC} \)
falls by several orders of magnitude\( ^{6} \). This indicates that the holes
are not fully mobile but delocalized over several lattice spacings. Inelastic
neutron scattering (INS) experiments reveal the existence of new states within
the Haldane gap. Several studies have been carried out so far to explain the
origin of the sub-gap states\( ^{9-14} \).

A very recent neutron scattering experiment provides evidence for an incommensurate double-peaked structure factor \( S(q) \) for the sub-gap states\( ^{15} \). The INS intensity is proportional to the structure factor. For the pure compound, \( S(q) \) near the gap energy of 9 mev has a single peak at the wave vector \( q=\pi  \), indicative of AFM correlations. For the doped compound, \( S(q) \) has an incommensurate double-peaked structure factor, for energy transfer \( \omega \sim  \)3-7 mev, with the peaks located at \( q=\pi \pm \delta q \). The shift \( \delta q \) is found to have a very weak dependence on the impurity concentration x for x in the range   x\( \in  \)[0.04 , 0.14]. Evidence of incommensurate peaks has also been obtained in the underdoped metallic cuprates\( ^{1} \). The peaks are four in number and occur at (\( \pi \pm \delta q,\pi  \)) and (\( \pi ,\pi \pm \delta q \) ). The crucial difference from the nickelate compound is that \( \delta q \) is proportional to the dopant concentration x. The incommensurability has been ascribed to inhomogeneous spin and charge ordering in the form of stripes. Recently, Malvezzi and Dagotto\( ^{16} \) have provided an explanation for the origin of spin incommensurability in the hole-doped S =1 nickelate compound. They have studied a two-orbital model using the Density Matrix Renormalization Group (DMRG) method. They have shown that a mobile hole generates AFM correlations between the spins located on both sides of the hole and this is responsible for the spin incommensurability seen in experiments. Xu et al\( ^{15} \) have given a different explanation for the origin of incommensurability. The holes doped into the QSL ground state of the S = 1 chain are located on the oxygen orbitals and carry spin. They induce an effective ferromagnetic (FM) interaction between the Ni spins on both sides. The incommensurate peaks arise because of the spin density modulations developed around the holes with the size of the droplets controlled by the correlation length of the QSL. In this paper, we provide a microscopic theory of the origin of spin incommensuration in keeping with the suggestions of Xu et al.

\section*{2. Microscopic model}

Various microscopic models have been proposed to explain the presence of sub-gap
states\( ^{9-14} \). The relevant orbitals of \( Ni^{2+} \) in the Ni-O chain
are \( 3d_{3z^{2}-r^{2}} \) and \( 3d_{x^{2}-y^{2}} \)\( ^{13} \). An electron
in the latter orbital is almost localised while the \( 3d_{3z^{2}-r^{2}} \)
orbital has a finite overlap with the \( 2p_{z} \) orbital of the oxygen ion.
In the undoped state, each Ni orbital is occupied by a single electron and the
oxygen orbital is filled up. The spins on the Ni orbitals have a FM alignment,
due to strong Hund's rule coupling, giving rise to total spin S = 1. The nearest-neighbour
(NN) spins S = 1 interact via AFM superexchange interaction mediated through
the intermediate oxygen ions. On doping, holes are predominantly introduced
in the oxygen orbitals. The holes have effective spin-1/2 as each \( 2p_{z} \)
orbital containing a hole is occupied by a single electron (Fig. 1). 

In the undoped state, we have a chain of S = 1 spins. The ground state is spin-disordered
and, as predicted by Haldane\( ^{8} \), there is a gap in the excitation spectrum.
Affleck, Kennedy, Lieb and Tasaki (AKLT)\( ^{17} \) have provided a physical
picture for the ground state. Each spin 1 can be considered to be a symmetric
combination of two spin-1/2's. Each spin-1/2 forms a singlet or valence bond
(VB), \( \frac{1}{\sqrt{2}}(\uparrow \downarrow -\downarrow \uparrow ), \)
with a NN spin-1/2. The ground state is described as a valence bond solid (VBS)
in which a singlet or VB is present between each NN pair of spins. Actually,
AKLT wrote down the Hamiltonian for which the VBS state is the exact ground
state. The Hamiltonian is a sum over projection operators onto spin 2 for successive
pairs of S = 1 spins and can be written as

\begin{equation}
\label{1}
H_{AKLT}=J\sum _{\left\langle ij\right\rangle }\left[ \overrightarrow{S_{i}}.\overrightarrow{S_{j}}+\frac{1}{3}(\overrightarrow{S_{i}}.\overrightarrow{S_{j}})^{2}+\frac{2}{3}\right] .
\end{equation}
 Since, in the VBS state, a singlet exists between two NN spin-1/2's, the total
spin of two NN spin-1's can never be 2. Thus, the projection operator onto spin
2 acting on a NN pair of spins in the VBS state gives zero. The VBS state is
the exact ground state of \( H_{AKLT} \) with eigenvalue zero. The VBS state
also provides the correct physical picture for the ground state of the S =1
chain interacting via the usual AFM Heisenberg Hamiltonian

\begin{equation}
\label{2}
H=J\sum ^{N}_{i=1}\overrightarrow{S}_{i}.\overrightarrow{S}_{i+1.}
\end{equation}
 This has been verified in a number of experiments.

Excitations from the VBS state are triplets (a VB is replaced by a triplet)
which propagate along the chain\( ^{18} \). The triplet excitation spectrum
is separated from the VBS state by the Haldane gap, of the order of J. The triplet
excitation spectrum has been observed experimentally in \( Y_{2}BaNiO_{5} \).
The excitation differs from a spin wave in that it propagates in a spin background
with no conventional magnetic order. The Haldane gap \( \Delta  \) has a magnitude
\( \Delta \simeq  \) 9 mev\( ^{15} \). The VBS state is not, however, a simple
paramagnet but has quantum coherence in the form of a long-range string order\( ^{19} \).
On doping the nickelate compound with holes, the triplet excitation spectrum
has a dispersion relation similar to that in the pure \( Y_{2}BaNiO_{5} \).
In addition, however, new states appear in the Haldane gap. Near 3 to 7 mev,
the magnetic scattering is found to be the most intense along two vertical lines,
displaced symmetrically from \( \pi  \). 

The oxygen hole acts as a spin-1/2 impurity and also modifies the strength of
the superexchange interaction between neighbouring Ni spins. Sorensen and Affleck\( ^{10} \)
have considered S = 1 spin chains with site/bond impurities to explain the origin of sub-gap states. The holes are, however, not localized but mobile over several lattice
spacings. Various models of fully mobile S = 1/2 holes interacting with S =
1 spins have been proposed so far\( ^{11,13,20,21} \). The first such model
is that of Penc and Shiba\( ^{11} \). The effective Hailtonian is given by

\begin{equation}
\label{3}
H=H_{0}+H_{J^{\prime }}+H_{\widetilde{t}}+H_{\widetilde{J}}
\end{equation}
 where 

\begin{equation}
\label{4}
H_{0}=J\sum _{i}\left[ \overrightarrow{S}_{i}.\overrightarrow{S}_{i+1}-\beta (\overrightarrow{S_{i}}.\overrightarrow{S}_{i+1})^{2}\right] .
\end{equation}
The special cases \( \beta  \) = 0 and \( \beta  \) = \( -\frac{1}{3} \)
correspond to the Heisenberg and AKLT Hamiltonians respectively. The Hamiltonian

\begin{equation}
\label{5}
H_{J^{\prime }}=2J^{\prime }\sum _{i}(\overrightarrow{S}_{i}.\overrightarrow{\sigma }_{i+1/2}+\overrightarrow{S}_{i+1}.\overrightarrow{\sigma }_{i+1/2})
\end{equation}
describes the effect of S = 1/2 impurities (holes). An impurity at site i+1/2
changes the J coupling to \( J_{1} \) between the S =1 spins on both sides
of the impurity in the Hamiltonian \( H_{0} \) (Eq.(4)). Also, there is an
interaction between the impurity spin \( \sigma  \) and the adjacent S = 1
spins. \( H_{\widetilde{t}} \) and \( H_{\widetilde{J}} \) are the Hamiltonians
describing the effective hopping of the holes on the O sites.

\begin{equation}
\label{6}
H_{\widetilde{t}}=\widetilde{t}\sum _{i}\widehat{P_{i}}
\end{equation}
and

\begin{equation}
\label{7}
H_{\widetilde{J}}=2\widetilde{J}\sum _{i}\widehat{P_{i}}(\overrightarrow{S}_{i}.\overrightarrow{\sigma }_{i+1/2}+\overrightarrow{\sigma }_{i-1/2}.\overrightarrow{S}_{i}).
\end{equation}
 The operator \( \widehat{P_{i}} \) exchanges the occupations of the sites
i +1/2 and i\( -1/2 \) . If the hole is located on site i+1/2, \( \widehat{P_{i}} \)
shifts it to i-1/2 provided that site is empty. Two S = 1/2 objects cannot occupy
the same site as it costs a lot of energy. The same type of effective Hamiltonian
has been used earlier to describe doped cuprate systems in a two-band (copper
and oxygen orbitals) scenario\( ^{22-24} \). In the cuprates too, the added
holes go to oxygen sites and each hole has an effective spin-1/2 which interacts
with the spin-1/2's of the \( Cu^{2+} \) ions on neighbouring sites. The major
differences between the nickelate and cuprate compounds are: \( Ni^{2+} \)
ion carries spin 1 as opposed to the spin-1/2 of the \( Cu^{2+} \) ion and
in the undoped state the latter compound has AFM LRO. 

For the doped cuprate systems, Aharony et al\( ^{25} \) have argued that the
holes generate an effective FM exchange interaction between the Cu ions on neighbouring
sites. Consider a pair of Cu ions and an intermediate hole. The interaction
is described by a Hamiltonian of the type given in Eq.(5) except that the spins
\( \overrightarrow{S_{i}} \) have a magnitude 1/2. The exchange interaction
strength \( J^{\prime } \) is expected to be larger than the interaction strength
J. The spins \( \overrightarrow{S_{i}} \) and \( \overrightarrow{S}_{i+1} \)
prefer to be parallel irrespective of the sign of \( J^{\prime } \). The spin-1/2
of the hole along with the two spin-1/2's of the adjacent Cu-ions constitute
a spin bubble or polaron. For the doped \( Y_{2}BaNiO_{5} \) compound, a similar
scenario holds true. The hole with a spin-1/2 provides a break in the VBS state
of the S =1 chain. The two Ni-ions which are NNs of the hole have unpaired spin-1/2's.
In the absence of the hole, these two spin-1/2's form a singlet in the VBS state.
In the presence of the hole, a spin bubble consisting of three spin-1/2's (the
central spin belonging to the oxygen hole) is obtained. The bubbles are embedded
in the original VBS state.

\section*{3. Structure factor S(q)}

The differential scattering cross-section in an INS experiment is proportional
to the dynamical structure factor \( S^{\alpha \alpha }(q,\omega ) \), (\( \alpha  \)
= x,y or z) given by\( ^{26} \)

\begin{eqnarray}
S^{\alpha \alpha }(q,\omega ) & = & \sum _{S^{\prime },M^{\prime }}\sum _{i,j}exp\{iq(r_{i}-r_{j})\}\left\langle SM\right| S^{\alpha }_{i}\left| S^{\prime }M^{\prime }\right\rangle \left\langle S^{\prime }M^{\prime }\right| S^{\beta }_{j}\left| SM\right\rangle \nonumber \\
 &  & \delta (\omega +E(S,M)-E(S^{\prime },M^{\prime })).\label{8} 
\end{eqnarray}
We assume a 1d spin system at T = 0. The magnetic ions are located at the sites
i and j with the associated magnetic form factors taken to be unity. The spin
state \( \left| SM\right\rangle  \) is the ground state where S is the total
spin of the state and M the z component of the total spin. The ground state
energy is given by E(S,M). The state \( \left| S^{\prime }M^{\prime }\right\rangle  \)
is an eigenstate of the system with energy \( E(S^{\prime },M^{\prime }) \).
The energy and momentum transfers in the scattering process are \( \omega  \)
and q respectively. 

For the doped \( Y_{2}BaNiO_{5} \) compound, the spin states are the VBS states
with spin bubbles nucleated around the doped holes. Let us consider the case
of a single static hole. The total spin of the state is equal to the total spin
of the bubble since the rest of the VBS state is a singlet. Since the bubble
consists of three spin-1/2's, the possible spin states correspond to total spin
S = 3/2 and 1/2 (two representations). The possible spin states \( \left| SM\right\rangle  \)
are (only the bubble spin configurations are shown):

\begin{eqnarray}
\left| \frac{3}{2},\frac{3}{2}\right\rangle  & = & \left| \uparrow \uparrow \uparrow \right\rangle \nonumber \\
\left| \frac{3}{2},\frac{1}{2}\right\rangle  & = & \frac{1}{\sqrt{3}}\left| \uparrow \downarrow \uparrow +\uparrow \uparrow \downarrow +\downarrow \uparrow \uparrow \right\rangle \nonumber \\
\left| \frac{3}{2},-\frac{1}{2}\right\rangle  & = & \frac{1}{\sqrt{3}}\left| \downarrow \uparrow \downarrow +\downarrow \downarrow \uparrow +\uparrow \downarrow \downarrow \right\rangle \nonumber \\
\left| \frac{3}{2},-\frac{3}{2}\right\rangle  & = & \left| \downarrow \downarrow \downarrow \right\rangle \label{9} 
\end{eqnarray}

\[
\left| \frac{1}{2},\frac{1}{2}\right\rangle =\frac{1}{\sqrt{6}}\left| 2\uparrow \downarrow \uparrow -\downarrow \uparrow \uparrow -\uparrow \uparrow \downarrow \right\rangle \]

\begin{equation}
\left| \frac{1}{2},-\frac{1}{2}\right\rangle =\frac{1}{\sqrt{6}}\left| 2\downarrow \uparrow \downarrow -\uparrow \downarrow \downarrow -\downarrow \downarrow \uparrow \right\rangle 
\end{equation}

\[
\left| \widetilde{\frac{1}{2}},\frac{1}{2}\right\rangle =\frac{1}{\sqrt{2}}\left| \uparrow \uparrow \downarrow -\downarrow \uparrow \uparrow \right\rangle \]

\begin{equation}
\left| \widetilde{\frac{1}{2}},-\frac{1}{2}\right\rangle =\frac{1}{\sqrt{2}}\left| \downarrow \downarrow \uparrow -\uparrow \downarrow \downarrow \right\rangle 
\end{equation}

The `\~{}'-sign in Eq.(11) indicates a different S = 1/2 representation. In
the states shown in Eqs.(9)-(10), the first and the third spins (the Ni spin-1/2's)
are in a S = 1 spin configuration. This spin 1 adds to the spin 1/2 of the hole
giving rise to a resultant spin 3/2 or 1/2. In all these states, there is thus
an effective FM interaction between the Ni spins. In the states shown in Eq.
(11), the Ni spin-1/2's form a singlet and the hole spin is free. 

Penc and Shiba\( ^{11} \) have calculated the hole excitation spectrum assuming
a fully mobile hole. The trial wave function of the L-site periodic chain with
one hole is given by

\begin{equation}
\label{12}
\left| SMk\right\rangle =\frac{1}{\sqrt{L}}\sum _{j}e^{ikj}\left| SM,j+\frac{1}{2}\right\rangle 
\end{equation}
where \( \left| SM,j+\frac{1}{2}\right\rangle  \) denotes the state with the
hole located at site \( j+\frac{1}{2} \). With \( H_{0} \) in Eq.(4) given
by the AKLT Hamiltonian, the lowest hole eigenstate of H (Eq.(3)) is given by
an appropriate linear combination of the states \( \left| \frac{1}{2}\frac{!}{2}k\right\rangle  \)
and \( \left| \widetilde{\frac{1}{2}}\frac{1}{2}k\right\rangle  \). An excited
state is obtained from a different linear combination of the states. The full
hole spectrum, in the subspace of states described by Eqs. (9)-(11), consists
of one fourfold and two twofold degenerate bands. The dynamical structure factor
has non-zero weight inside the Haldane gap due to scattering between the hole
states. Penc and Shiba's microscopic model Hamiltonian has been rederived by
Batista et al\( ^{14} \) starting from a multiband Hamiltonian containing the
relevant Ni and O orbitals. They have further estimated the magnitudes of the
various parameters in the Hamiltonian on the basis of small cluster calculations. 

We return to our consideration of a static hole. From Eq.(8), the selection
rules for INS transitions are

\begin{equation}
\label{13}
\Delta S=0,\pm 1,\Delta M=0,\pm 1.
\end{equation}
 We consider \( \alpha  \) = z so that \( \Delta M=0 \) . The q-dependence
of the INS intensity is given by

\begin{equation}
\label{14}
I_{S^{\prime }M^{\prime }}(q)=\left| \left\langle S^{\prime }M^{\prime }\right| \sum _{i}e^{iqr_{i}}S^{z}_{i}\left| SM\right\rangle \right| ^{2}.
\end{equation}
 \( I_{S^{\prime }M^{\prime }}(q) \) corresponds to a scattering transition
from the state \( \left| SM\right\rangle  \) to the state \( \left| S^{\prime }M^{\prime }\right\rangle  \)
with an energy transfer of \( \omega  \). The spin expectation values in Eq.(14)
can be calculated in a straightforward manner using the method of matrix products\( ^{27-30} \).
In the undoped state, the Hamiltonian H in Eq.(3) reduces to \( H_{0} \) .
We consider \( H_{0} \) to be the AKLT Hamiltonian for which the VBS state
is the exact ground state. In the matrix product representation, the VBS state
is given by

\begin{equation}
\label{15}
\left| \psi _{VBS}\right\rangle =Tr(g_{-L}\otimes ....\otimes g_{0}\otimes ....\otimes g_{L}).
\end{equation}
 The 1d lattice consists of 2L + 1 sites and \( g_{i} \) denotes a \( 2\times 2 \)
matrix at the ith site with single spin states as elements.

\begin{equation}
\label{16}
g=\frac{1}{\sqrt{3}}\left( \begin{array}{cc}
-\left| 0\right\rangle  & -\sqrt{2}\left| 1\right\rangle \\
\sqrt{2}\left| \overline{1}\right\rangle  & \left| 0\right\rangle 
\end{array}\right) 
\end{equation}
 where \( \left| 1\right\rangle ,\left| 0\right\rangle ,\left| \overline{1}\right\rangle  \)
are the spin-1 states with \( S^{z}=+1,0 \) and \( -1 \) respectively. Expectation
values in the matrix product states can be calculated most conveniently by the
transfer-matrix method. We describe the major steps of the calculation in the
following. Details of the theory of matrix product representations can be obtained
from Refs. {[}27-30{]}.

The norm of a wave function in the matrix product representation, e.g., \( \left\langle \psi _{VBS}\mid \psi _{VBS}\right\rangle  \)
can be written as

\begin{equation}
\label{17}
\left\langle \psi _{VBS}\mid \psi _{VBS}\right\rangle =\sum _{\{n_{\alpha },m_{\alpha }\}}g^{\dagger }_{n_{1}n_{2}}g^{\dagger }_{n_{2}n_{3}}........g^{\dagger }_{n_{L}n_{1}}g_{m_{1}m_{2}}g_{m_{2}m_{3}}.....g_{m_{L}m_{1}.}
\end{equation}
 We define a \( 4\times 4 \) transfer matrix G at any lattice site by

\begin{equation}
\label{18}
G_{\mu _{1}\mu _{2}}\Rightarrow G_{(n_{1}m_{1}),(n_{2}m_{2})}=g^{\dagger }_{n_{1}n_{2}}g_{m_{1}m_{2}}.
\end{equation}
 The ordering of multi-indices is given by

\begin{equation}
\label{19}
\mu =1,2,3,4\leftrightarrow (11),(12),(21),(22).
\end{equation}
 With the choice of g given in Eq.(16), the transfer matrix G is

\begin{equation}
\label{20}
G=\frac{1}{3}\left( \begin{array}{cccc}
1 & 0 & 0 & 2\\
0 & -1 & 0 & 0\\
0 & 0 & -1 & 0\\
2 & 0 & 0 & 1
\end{array}\right) .
\end{equation}
 The eigenvalues and eigenvectors of G are

\begin{equation}
\label{21}
\lambda _{1}=1,\lambda _{2}=\lambda _{3}=\lambda _{4}=-\frac{1}{3}
\end{equation}
 
\begin{equation}
\label{22}
\left| e_{1}\right\rangle =\frac{1}{\sqrt{2}}\left( \begin{array}{c}
1\\
0\\
0\\
1
\end{array}\right) ,\left| e_{2}\right\rangle =\frac{1}{\sqrt{2}}\left( \begin{array}{c}
-1\\
0\\
0\\
1
\end{array}\right) ,\left| e_{3}\right\rangle =\left( \begin{array}{c}
0\\
1\\
0\\
0
\end{array}\right) ,\left| e_{4}\right\rangle =\left( \begin{array}{c}
0\\
0\\
1\\
0
\end{array}\right) .
\end{equation}
The normalisation (17) is determined by the largest eigenvalue for \( L\rightarrow \infty  \)
(the size of the chain is 2L+1):

\begin{equation}
\label{23}
\left\langle \psi _{VBS}\mid \psi _{VBS}\right\rangle \simeq \lambda ^{2L+1}_{1}=1.
\end{equation}
Consider the ground state matrix element of a spin operator P at site r. One
uses the notation

\begin{equation}
\label{24}
Z(P)_{\mu _{1}\mu _{2}}\Rightarrow Z(P)_{(n_{1}m_{1}),(n_{2}m_{2})}=g^{\dagger }_{n_{1}n_{2}}Pg_{m_{1}m_{2}}
\end{equation}
 with Z(1)=G. We obtain

\begin{equation}
\label{25}
\left\langle P\right\rangle =\left\langle \psi _{VBS}\mid \psi _{VBS}\right\rangle ^{-1}\left\langle \psi _{VBS}\mid P\mid \psi _{VBS}\right\rangle =(TrG^{2L+1})^{-1}TrZ(P)G^{2L}.
\end{equation}
 On taking the limit \( L\rightarrow \infty  \) , we obtain

\begin{equation}
\label{26}
\left\langle P\right\rangle =\lambda ^{-1}_{1}\left\langle e_{1}\right| Z(P)\left| e_{1}\right\rangle .
\end{equation}
 For 2-site correlations of operators \( A_{1} \) and \( B_{r} \) at sites
1 and r respectively, one obtains

\begin{equation}
\label{27}
\left\langle A_{1}B_{r}\right\rangle =Tr(G^{2L+1})^{-1}TrZ(A)G^{r-2}Z(B)G^{2L+1-r}.
\end{equation}
 For \( L\rightarrow \infty  \), the expectation value reduces to

\begin{equation}
\label{28}
\left\langle A_{1}B_{r}\right\rangle =\sum ^{4}_{n=1}\lambda ^{-2}_{n}\left( \frac{\lambda _{n}}{\lambda _{1}}\right) ^{r}\left\langle e_{1}\right| Z(A)\left| e_{n}\right\rangle \left\langle e_{n}\right| Z(B)\left| e_{1}\right\rangle 
\end{equation}
 where \( \lambda _{n} \) and \( \left| e_{n}\right\rangle  \) are the eigenvalues
and eigenvectors of the transfer matrix G. In the same way, one can calculate
off-diagonal expectation values of the type \( \left\langle S^{\prime }M^{\prime }\right| A_{i}\left| SM\right\rangle  \).

The VBS state with a single embedded hole has different matrix product representations
depending on the configuration of the three-spin bubble centred around the hole.
The possible bubble spin configurations are given in Eqs. (9)-(11). For example,
the states \( \left| \frac{3}{2}\frac{3}{2}\right\rangle  \) and \( \left| \frac{1}{2}\frac{1}{2}\right\rangle  \)
are of the form

\begin{equation}
\label{29}
Tr(g_{-L}\otimes .....\otimes g_{-1}\otimes g^{h}\otimes g_{0}\otimes ....\otimes g_{L})
\end{equation}
 with the hole located in between the sites -1 and 0. The corresponding matrix
\( g^{h} \) is given by

\begin{equation}
\label{30}
g_{h}=\left( \begin{array}{cc}
0 & 0\\
-\sqrt{2}\left| \uparrow \right\rangle  & 0
\end{array}\right) 
\end{equation}
 for the \( \left| \frac{3}{2}\frac{3}{2}\right\rangle  \) state and

\begin{equation}
\label{31}
g_{h}=\frac{1}{\sqrt{3}}\left( \begin{array}{cc}
-\left| \uparrow \right\rangle  & 0\\
-2\left| \downarrow \right\rangle  & \left| \uparrow \right\rangle 
\end{array}\right) 
\end{equation}
for the \( \left| \frac{1}{2}\frac{1}{2}\right\rangle  \) state. 

We have calculated \( I_{S^{\prime }M^{\prime }}(q) \) in Eq.(14) by taking
the ground state \( \left| SM\right\rangle  \) to be the \( \left| \frac{1}{2}\frac{1}{2}\right\rangle  \)
state and the state \( \left| S^{\prime }M^{\prime }\right\rangle  \) to be
a state which satifies the selection rule \( \Delta S=0,\pm 1 \) and \( \Delta M=0 \)
. For example, the state \( \left| S^{\prime }M^{\prime }\right\rangle  \)
can be any one of the states \( \left| \frac{3}{2}\frac{1}{2}\right\rangle  \),
\( \left| \frac{1}{2}\frac{1}{2}\right\rangle  \) and \( \left| \widetilde{\frac{1}{2}}\frac{1}{2}\right\rangle  \).
In the first two cases, \( I_{S^{\prime }M^{\prime }}(q) \) is calculated as
(apart from numerical prefactors )

\begin{equation}
\label{32}
I_{S^{\prime }M^{\prime }}(q)=\left| \frac{(1+e^{K})cos(\frac{q}{2})}{coshK+cosq}\right| ^{2}
\end{equation}
 where \( K^{-1}=\frac{1}{ln3} \) is the spin correlation length in the VBS
state. For the state \( \left| \frac{3}{2}\frac{1}{2}\right\rangle  \) ,

\begin{equation}
\label{33}
g_{h}=\sqrt{\frac{2}{3}}\left( \begin{array}{cc}
\left| \uparrow \right\rangle  & 0\\
-\left| \downarrow \right\rangle  & -\left| \uparrow \right\rangle 
\end{array}\right) 
\end{equation}
Also, i , j in Eq.(8) denote the sites at which the Ni spins are located, i.e.,
the hole spin is not explicitly taken into account. When \( \left| S^{\prime }M^{\prime }\right\rangle  \)
is \( \left| \widetilde{\frac{1}{2}}\frac{1}{2}\right\rangle  \) ,

\begin{equation}
\label{34}
g_{h}=\left( \begin{array}{cc}
\left| \uparrow \right\rangle  & 0\\
0 & \left| \uparrow \right\rangle 
\end{array}\right) 
\end{equation}
and \( I_{S^{\prime }M^{\prime }}(q) \) is given by (apart from a numerical
prefactor),

\begin{equation}
\label{35}
I_{S^{\prime }M^{\prime }}(q)=\left| \frac{(1+e^{K})sin(\frac{q}{2})}{coshK+cosq}\right| ^{2}
\end{equation}
Xu et al\( ^{15} \) have suggested a form (Eq.(32)) for the structure factor
of the ground state with a single FM bond inserted within an infinite chain.
The FM bond occurs between the sites -1 and 0. In the matrix product representation,
such a state is given by

\begin{equation}
\label{36}
\left| \psi _{FM}\right\rangle =Tr(g_{-L}\otimes .....\otimes g^{a}_{-1}\otimes g_{0}\otimes ....\otimes g_{L})
\end{equation}

where

\begin{equation}
\label{37}
g^{a}=\left( \begin{array}{cc}
\sqrt{2}\left| 1\right\rangle  & 0\\
-\left| 0\right\rangle  & 0
\end{array}\right) 
\end{equation}

Calculation of S(q) with \( \left| S^{\prime }M^{\prime }\right\rangle =\left| SM\right\rangle =\left| \psi _{FM}\right\rangle  \)
reproduces the form in Eq.(32). This, however, corresponds to elastic neutron
scattering whereas the incommensurate signal arises from inelastic neutron scattering.
The intensity \( I_{S^{\prime }M^{\prime }}(q) \) (Eq.(32)) has two peaks symmetrically
displaced from \( \pi  \). The widths and incommensurability are of the order
of K. A good fit to the experimental data has been obtained by treating \( K^{-1} \)
as a parameter. For dopant concentrations x = 0.04, 0.095 and 0.14, the best
fits to the experimental data are obtained for \( K^{-1}=8.1\pm 0.2,7.3\pm 0.2 \)
and \( 7.2\pm 0.5 \) respectively. These values are considerably higher than
the VBS value of \( K^{-1}=\frac{1}{ln3} \) and are closer to the estimate
of the correlation length for the Heisenberg Hamiltonian. The use of the AKLT
Hamiltonian has, however, enabled us to formulate a microscopic theory of the
structure factor, based on the matrix product representation. The double-peaked
incommensurate structure is correctly reproduced in the theory. The intensity
\( I_{S^{\prime }M^{\prime }}(q) \) , however, has nodes at \( q=(2n+1)\pi  \)
which are not seen experimentally. Xu et al\( ^{15} \) have suggested that
finite hole densities as well as an explicit consideration of the hole spin
distributed over l lattice sites centred on the FM bond, gives rise to a non-zero
intensity at \( q=(2n+1)\pi  \). Microscopic calculations along these lines
are, however, yet to be carried out.

The structure factor S(q) of INS is obtained on integrating the dynamical structure
factor \( S(q,\omega ) \) in Eq.(8) over energy \( \omega  \) and is given
by the sum over the intensities \( I_{S^{\prime }M^{\prime }}(q) \)'s for the
three possible states \( \left| S^{\prime }M^{\prime }\right\rangle  \). Two
of the intensities is proportional to \( cos^{2}\frac{q}{2} \) while the third
is proportional to \( sin^{2}\frac{q}{2} \). The first two intensities have
a minimum at \( q=\pi  \) and the last intensity has a maximum at \( q=\pi  \).
The resultant intensity for \( K=ln3\sim 1.1 \) has a double-peaked incommensurate
structure but the intensity at \( q=\pi  \) has a magnitude very close to that
of the peak intensities. For K = 1.2, (correlation length less than NN Ni-Ni
distance but more than the distance separating NN Ni and oxygen ions), the structure
factor S(q) vs. q is shown in Fig.2. The double-peaked structure is, however,
very sensitive to the value of K and disappears for \( K\leq 1.0 \). For exchange
interactions of the Heisenberg-type, K is around 0.17. With a ground state \( \left| SM\right\rangle =\left| \frac{1}{2}\frac{1}{2}\right\rangle +x\left| \widetilde{\frac{1}{2}}\frac{1}{2}\right\rangle  \)
(x < 1), the structure factor S(q) contains terms proportional to \( cos^{2}\frac{q}{2} \)
as well as \( \delta sin^{2}\frac{q}{2} \) (\( \delta  \) < 1). With the appropriate
choice of \( \delta  \), a double-peaked incommensurate structure factor S(q)
is obtained for a wide range of values of K. Fig.3 shows S(q) vs. q for \( \delta =0.35 \)
and K = ln3. Eigenfunctions which are linear combinations of the states \( \left| \frac{1}{2}\frac{1}{2}\right\rangle  \)
and \( \left| \widetilde{\frac{1}{2}}\frac{1}{2}\right\rangle  \) are obtained
when mobile holes are considered. In fact, as mentioned earlier, for fully mobile
holes , the lowest two hole bands correspond to linear combinations of the states
\( \left| \frac{1}{2}\frac{1}{2}k\right\rangle  \) and \( \left| \widetilde{\frac{1}{2}}\frac{1}{2}k\right\rangle  \)
with \( \left| SMk\right\rangle  \) defined as in Eq.(18). Holes in the doped
nickelate compound are not, however, fully mobile. Calculations, taking into
account the limited mobility of a hole as well as an explicit consideration
of the hole spin in calculating S(q), are in progress and the results will be
reported elsewhere.

\section*{4. Conclusion}

In this paper, we have provided a microscopic theory for the double-peaked structure
of the INS intensity for the doped S = 1 AFM compound \( Y_{2}BaNiO_{5} \).
The spin-1's are assumed to interact via the AKLT Hamiltonian for which the
VBS state is known to be the exact ground state. The introduction of holes in
the system nucleates spin bubbles around the holes due to the strong AFM exchange
interactions between the hole spins and NN free spin-1/2's of the VBS state.
The scattering between the hole states gives rise to states within the Haldane
gap with an incommensurate structure factor. The incommensurate peaks occurring
in the underdoped metallic cuprate systems have been ascribed to the formation
of spin and charge inhomogeneities in the form of stripes. As a consequence
of this, the shift \( \delta q \) of the location of an incommensurate peak
from \( \pi  \) is proportional to the dopant concentration x. For the doped
nickelate system, \( \delta q \) has a very weak dependence on x for x in the
range 0.04-0.14 and no experimental evidence of stripes has been obtained. Incommensuration
in the doped cuprate systems occurs in the metallic state when holes are fully
mobile. In the doped cuprates, the spin-spin correlations have a longer range
than in the case of the doped nickelate system. This gives rise to greater effective
interactions between the spin bubbles in the former case.

Malvezzi and Dagotto\( ^{16} \) have considered a two-orbital model for the
doped \( Y_{2}BaNiO_{5} \) compound and have shown that spin incommensurability
is obtained in a wide range of densities and couplings. AFM correlations are
dynamically generated across the holes to facilitate hole movement. The same
type of correlations occur in the 1d Hubbard model and the 2d extended t-J model.
In the two-orbital model, only the two Ni orbitals have been considered and
oxygen orbitals have not been explicitly included in the Hamiltonian. The spin
of the hole on the oxygen ion forms a singlet with the spin-1/2 in the adjacent
Ni orbital (each of the two Ni orbitals contains an electron with spin-1/2)
giving rise to an effective Zhang-Rice (ZR) doublet. This is analogous to a
ZR singlet in the case of doped cuprate systems\( ^{2} \). More detailed experiments
are required to distinguish between the two mechanisms, suggested so far, on
the origin of spin incommensuration in \( Y_{2-x}Ca_{x}BaNiO_{5} \). 

Several earlier studies on doped cuprate systems have explored the possibility
of the binding of two spin bubbles in the antiferromagnetically interacting
spin background. Zhang and Arovas\( ^{31} \) have considered the propagation
of S = 0 holes in a VBS state and have found evidence for the binding of holes.
The propagating S = 0 hole corresponds to a simultaneous hopping of a pair of
electrons. The possibility for the binding of a pair of spin bubbles, centred
around holes, in a VBS-type spin background has not been explored as yet. Experimental
evidence of the binding of holes in the doped nickelate compound does not exist.
On the other hand, evidence for the binding of holes has been obtained in the
case of a doped ladder compound\( ^{3,5} \). Several numerical studies provide
evidence for the binding of holes in a two-chain t-J ladder model\( ^{3,32} \).
The binding of two holes has been demonstrated exactly and analytically in a
specially constructed t-J ladder model\( ^{33} \). In the undoped case, this
model, in an appropriate parameter regime, has a ground state identical to that
of a S = 1 AFM Heisenberg spin chain\( ^{34} \). The exact results on the binding
of two holes hold true in a different parameter regime. The connection between
doped spin-1 and ladder systems is not clearly established as yet. In summary,
the experimental results on the S = 1 doped nickelate compound provide examples
of the rich phenomena observed in other spin systems. Comparative studies of
such systems are needed to highlight the common origins, if any, of some of
these phenomena.

\newpage

\section*{Figure Captions}

\textbf{Fig. 1.} Schematic diagram of the Ni-O chain with the second O being
occupied by a single hole.\\
\textbf{Fig. 2.} Structure factor S(q) vs. q for K = 1.2\\
\textbf{Fig. 3.} Structure factor S(q) vs. q for K = ln3 and \( \delta  \)
= 0.35

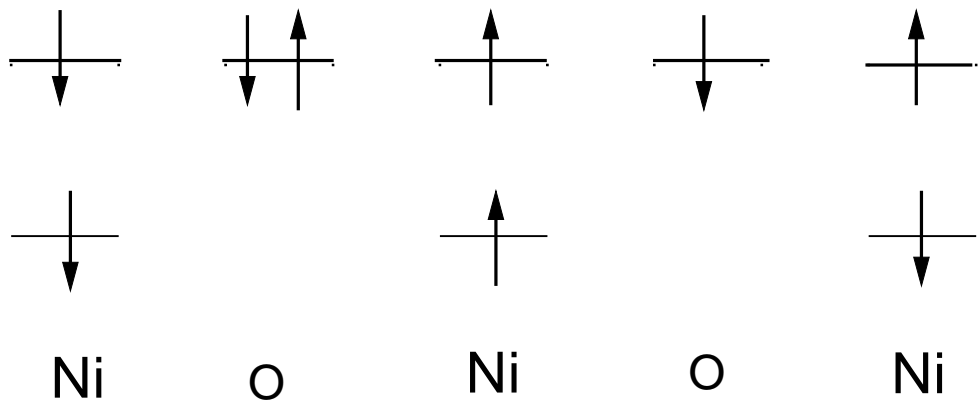
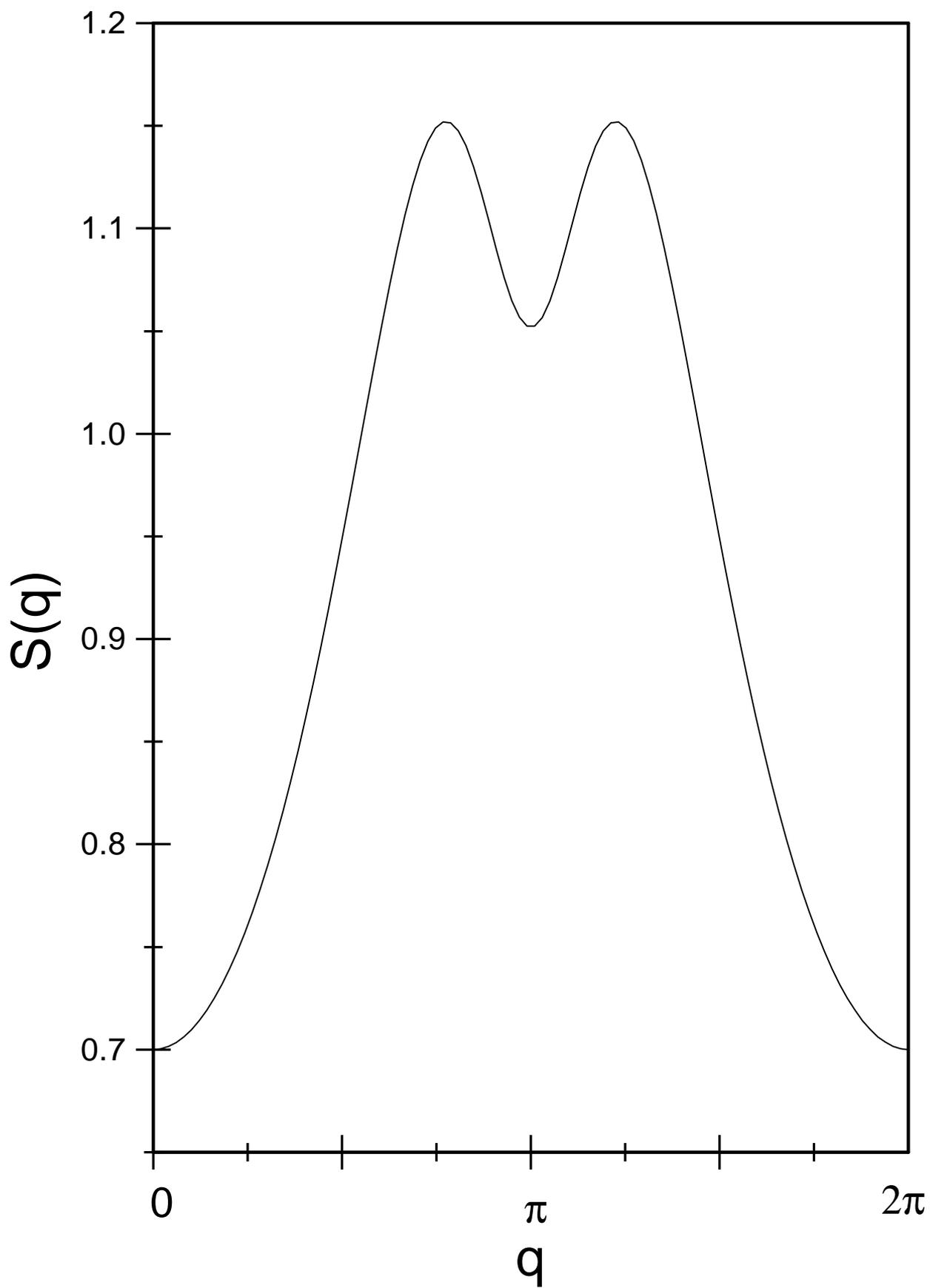
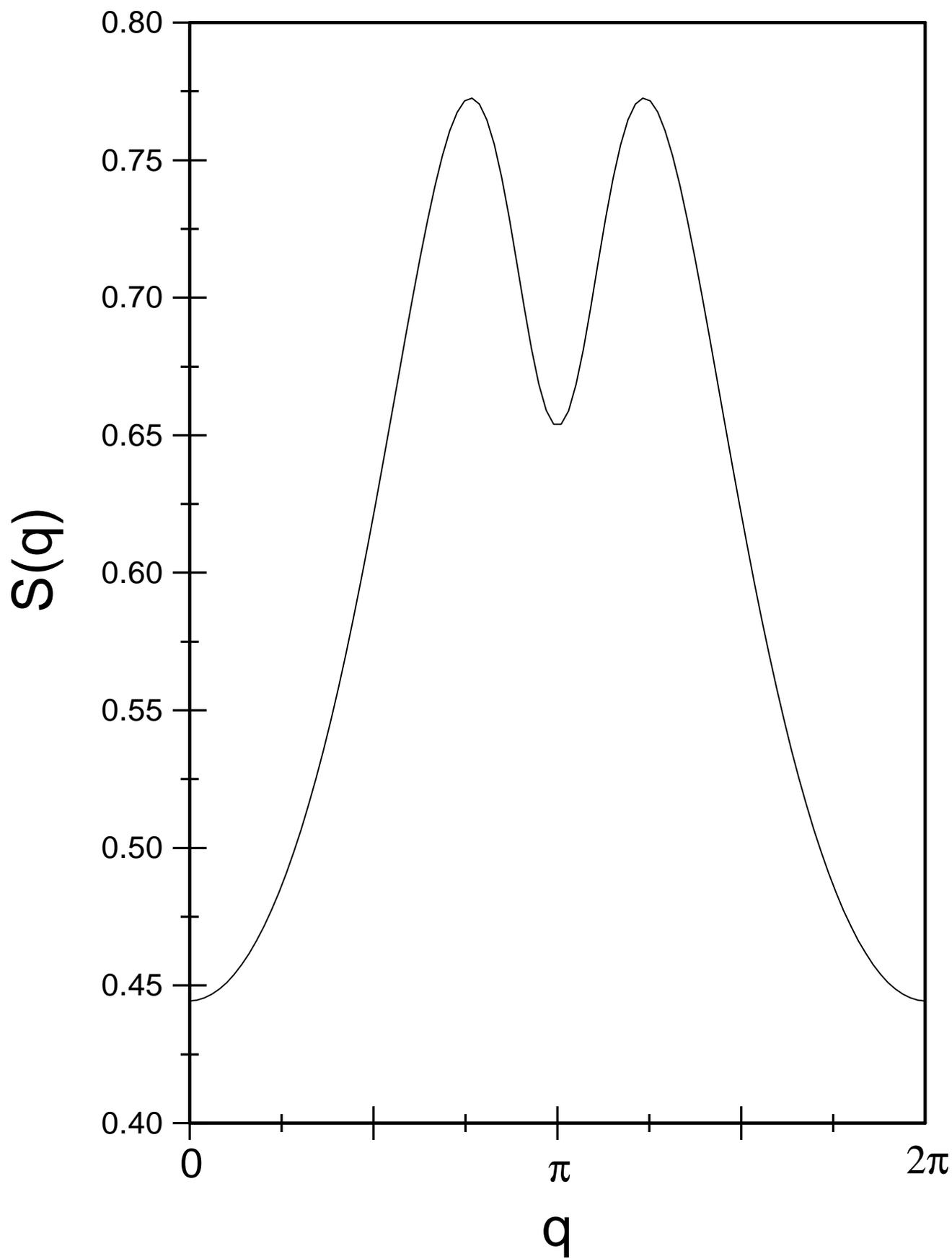
\end{document}